\documentclass[aps,prl,superscriptaddress,showkeys,showpacs,twocolumn,longbibliography,nofootinbib]{revtex4-1}
\usepackage[utf8]{inputenc}
\usepackage{amsmath}
\usepackage{amsfonts}
\usepackage{amsthm}
\usepackage{float}
\usepackage{braket}
\usepackage{graphicx}
\usepackage{url}
\usepackage[colorlinks=true, pdfstartview=FitV, linkcolor=red, citecolor=blue, urlcolor=blue]{hyperref}
\usepackage{slashed}
\usepackage[normalem]{ulem}

\graphicspath{{./figures/}}

\newcommand{\be}{\begin{equation}}      
\newcommand{\ee}{\end{equation}}      
\newcommand{\bea}{\begin{eqnarray}}      
\newcommand{\eea}{\end{eqnarray}}

\begin{document}

\title{Non-linear chiral magnetic waves}

\author{Kazuki Ikeda}
\email[]{kazuki.ikeda@stonybrook.edu}
\affiliation{Co-design Center for Quantum Advantage, Stony Brook University, Stony Brook, New York 11794-3800, USA}
\affiliation{Center for Nuclear Theory, Department of Physics and Astronomy, Stony Brook University, Stony Brook, New York 11794-3800, USA}

\author{Dmitri E. Kharzeev}
\email[]{dmitri.kharzeev@stonybrook.edu}
\affiliation{Center for Nuclear Theory, Department of Physics and Astronomy, Stony Brook University, Stony Brook, New York 11794-3800, USA}
\affiliation{Department of Physics, Brookhaven National Laboratory, Upton, New York 11973-5000, USA}

\author{Shuzhe Shi}
\email[]{shuzhe-shi@tsinghua.edu.cn}
\affiliation{Center for Nuclear Theory, Department of Physics and Astronomy, Stony Brook University, Stony Brook, New York 11794-3800, USA}
\affiliation{Department of Physics, Tsinghua University, Beijing 100084, China}

\bibliographystyle{unsrt}

\begin{abstract}
The chiral magnetic wave (CMW) is a macroscopic quantum phenomenon that arises due to the mixing of the electric and chiral charge oscillations induced by the chiral anomaly. In this study we report the first quantum simulation (on classical hardware) of the real-time dynamics of CMWs in Schwinger model. Our quench protocol is the following: at $t=0$ we suddenly place an electric dipole at the middle of our lattice. Due to chiral anomaly, this dipole excites the CMW that propagates towards the edges of the lattice.  
In Schwinger model tuned to the conformal critical point (at $\theta = \pi$, $m/g \simeq 0.2$), we find a gapless linear CMW that propagates with the speed of light. For massless Schwinger model ($\theta =0, m=0$), we find a gapped linear CMW, in accord with previous analytical analyses. For massive Schwinger model (that is dual to strongly interacting bosonic theory), we enter the new regime of nonlinear CMWs, where we find a surprise. Specifically, for $m/g > 1$, the frequency of electric charge oscillations becomes much smaller than the frequency of the oscillations of the chiral charge. For $m/g =4$, we find a solution corresponding to a nearly static electric dipole with fast oscillations of the chiral charge confined within. We call this solution a ``thumper" and study its properties in detail. 
\end{abstract}

\maketitle

\emph{Introduction}. --- 
In the presence of an external magnetic field, the chirality imbalance (i.e. the difference in the densities of right- and left-handed fermions) induces an electric current directed along the direction of magnetic field -- this is the chiral magnetic effect (CME) \cite{Kharzeev:2004ey, Kharzeev:2007jp, Fukushima:2008xe}; for reviews, see \cite{Kharzeev:2013ffa,Kharzeev:2015znc,Landsteiner:2016led,Kharzeev:2020jxw}. CME is an inherently non-equilibrium phenomenon stemming from the non-conservation of chiral charge dictated by the chiral anomaly. The non-equilibrium nature of CME becomes particularly apparent in the emergence of a collective chiral excitation stemming from the anomaly-induced coupling of chiral charge oscillations to the oscillations of electric charge -- the chiral magnetic wave (CMW) \cite{Kharzeev:2010gd}. 

It is instructive to investigate the physics of CMW in $(1+1)$ dimensional models, corresponding to the strong magnetic field limit of $(3+1)$ theories. In particular, the case of $(1+1)$ dimensional massless QED (Schwinger model) has been discussed already in \cite{Kharzeev:2010gd}. As is well known, massless Schwinger model is exactly solvable by bosonization, and its free massive bosonic excitation can be interpreted as a result of mixing between the gauge field and the CMW \cite{Kharzeev:2010gd}. 

In this paper, we will extend the analysis of the CMW to the massive Schwinger model. This model is interesting because it possesses confinement, and in bosonic representation describes non-linear strongly interacting theory. Since the massive Schwinger model is not solvable analytically, and we are interested in the real-time, non-equilibrium behavior, we will rely on quantum simulations (on classical hardware). Many interesting aspects of 
 (1+1)-dimensional quantum field theories have been successfully addressed using quantum simulations, see~\cite{Klco:2018kyo,Butt:2019uul,Magnifico:2019kyj,Shaw:2020udc,Kharzeev:2020kgc,Ikeda:2020agk,2023arXiv230111991F,PhysRevD.107.L071502,Ikeda:2023zil} for examples and~\cite{Bauer:2022hpo} for a recent review of quantum simulation approach. 

\emph{Schwinger model}. --- 
The Lagrangian density of the Schwinger model~\cite{Schwinger:1962tp} is 
\begin{equation}
\label{eq:L0}
    \mathcal{L} = -\frac{1}{4}F_{\mu\nu}F^{\mu\nu}+\bar{\psi}(i\gamma^\mu\partial_\mu-g\gamma^\mu A_\mu-m)\psi. 
\end{equation}
We label the space-time coordinate by $x^\mu=(t,z)$. We denote the Pauli matrices as $X$, $Y$, and $Z$, and use the following convention for the Dirac matrices: $\gamma^0 = Z$, $\gamma^1 = i\,Y$, $\gamma^5=\gamma^0 \gamma^1 = X$.
In $(1+1)$ dimensions, the axial charge density $q_5(x)\equiv\bar{\psi}\gamma^5\gamma^0\psi(x)$ 
and the vector current density $j(x)\equiv\bar{\psi}\gamma^1\psi(x)$ are related 
by $q_5(x) = -j(x)$. Likewise, the vector charge density $q(x)\equiv\bar{\psi}\gamma^0\psi(x)$ and the 
axial current density $j_5(x)\equiv\bar{\psi}\gamma^5\gamma^1\psi(x)$ are related by $q(x) = j_5(x)$.

Because of these relations, the conservation of vector charge can be expressed as
\begin{align}
    &\partial_t q - \partial_z q_5 = 0.
    \label{eq:charge_conservation}
\end{align}
To obtain the equation for the CMW, we combine this relation with the conservation of axial charge $\partial_\mu J_5^{\mu} = 2i \, m \bar\psi \gamma^5 \psi$ (in the absence of an external electric field), and use the bosonization dictionary, in which  $q \to -\frac{\partial_z \phi}{\sqrt{\pi}}$, $q_5 \to -\frac{\partial_t \phi}{\sqrt{\pi}}$, and $i\bar\psi \gamma^5 \psi \to - c\, M \sin(2\sqrt{\pi}\phi)$, with $M=g/\sqrt{\pi}$ and $c=e^\gamma/(2\pi)$. The resulting equation describing the CMW in the Schwinger model then reads 
\begin{align}
    (\partial_t^2 - \partial_z^2+ M^2) \phi + 2\sqrt{\pi} c\,m\,M \sin(2\sqrt{\pi}\phi) = 0.
    \label{eq:classical_cmw}
\end{align}
It is clear that for the massless case $m=0$, the equation is linear and describes the propagation of a gapped excitation with mass $M=g/\sqrt{\pi}$ -- this is a familiar non-interacting bosonic representation of the Schwinger model. The dispersion relation of this bosonic excitation can be derived as a result of mixing between the gapless CMW and a plasmon mode, see \cite{Kharzeev:2010gd}. For massive case $m\neq0$, the CMW equation becomes nonlinear. The case of $\theta = \pi$ can be obtained by flipping the sign of fermion mass, $m \to -m$. Near the critical point at $\theta=\pi$ and $m/g \simeq 0.3$ (in the continuum case), the potential in \eqref{eq:classical_cmw} becomes nearly flat, and the dynamics becomes close to conformal, see \cite{Ikeda:2023zil}. In this case, we expect to see a gapless linear CMW.

\emph{The lattice Hamiltonian} --- 
To discretize our Hamiltonian, we use staggered fermions~\cite{Kogut:1974ag, Susskind:1976jm}
\begin{equation}
    \psi_1(x)\to\frac{\chi_{2n}}{\sqrt{a}},~\psi_2(x)\to\frac{\chi_{2n+1}}{\sqrt{a}},
\end{equation}
where $a$ is the finite lattice spacing. Then the lattice Hamiltonian corresponding to Eq.~\eqref{eq:L0} is
\begin{align}
\begin{aligned}
\label{eq:lattice_total_ham3}
H=&-\frac{i}{2a}\sum_{n=1}^{N-1}\big[\chi^\dagger_{n+1}\chi_{n}-\chi^\dagger_{n}\chi_{n+1}\big]\\
&+m\sum_{n=1}^{N}(-1)^n\chi^\dagger_n\chi_n+\frac{ag^2}{2}\sum_{n=1}^{N-1}L^2_n,
\end{aligned}
\end{align}
where $L_n$ is the electric field operator satisfying the Gauss' law constraint
\begin{equation}
    L_{n}-L_{n-1} =  \chi_n^\dagger \chi_n-\frac{1-(-1)^n}{2}.
    \label{eq:gauss_staggered}
\end{equation}

For the purpose of quantum simulation, let us put the lattice Hamiltonian in the spin representation using the Jordan--Wigner transformation~\cite{Jordan:1928wi}:
\begin{align}
\begin{aligned}
 \chi_n=\frac{X_n-iY_n}{2}\prod_{i=1}^{n-1}(-i Z_i).
\end{aligned}
\end{align}
The Hamiltonian of the model then becomes
\begin{align}
\begin{aligned}
\label{eq:Ham}
H=&
     \frac{1}{4a}\sum_{n=1}^{N-1}\Big[X_n X_{n+1}+Y_n Y_{n+1}\Big]
\\&
     +\frac{m}{2}\sum_{n=1}^N(-1)^n Z_n+\frac{a g^2}{2}\sum_{n=1}^{N-1}L^2_n
 \end{aligned}
 \end{align}
and the local vector and axial charge densities are, respectively, 
\begin{align}
    Q_n \equiv \,& \bar{\psi}\gamma^0\psi = \frac{Z_n+(-1)^n}{2a},\\
    Q_{5,n} \equiv \,& \bar{\psi}\gamma^5\gamma^0\psi = \frac{X_nY_{n+1}-Y_nX_{n+1}}{4a}\,.
\end{align}
For later convenience, we define the total charge operator $Q \equiv a\sum_{n=1}^{N} Q_n$, which commutes with the Hamiltonian.
With the boundary condition $L_0=0$, the Gauss' law constraint~\eqref{eq:gauss_staggered} leads to the solution
\begin{align}
    L_n =  a\sum_{j=1}^n Q_j\,.
\end{align}

\begin{figure}[!htbp]\centering
    \includegraphics[width=0.45\textwidth]{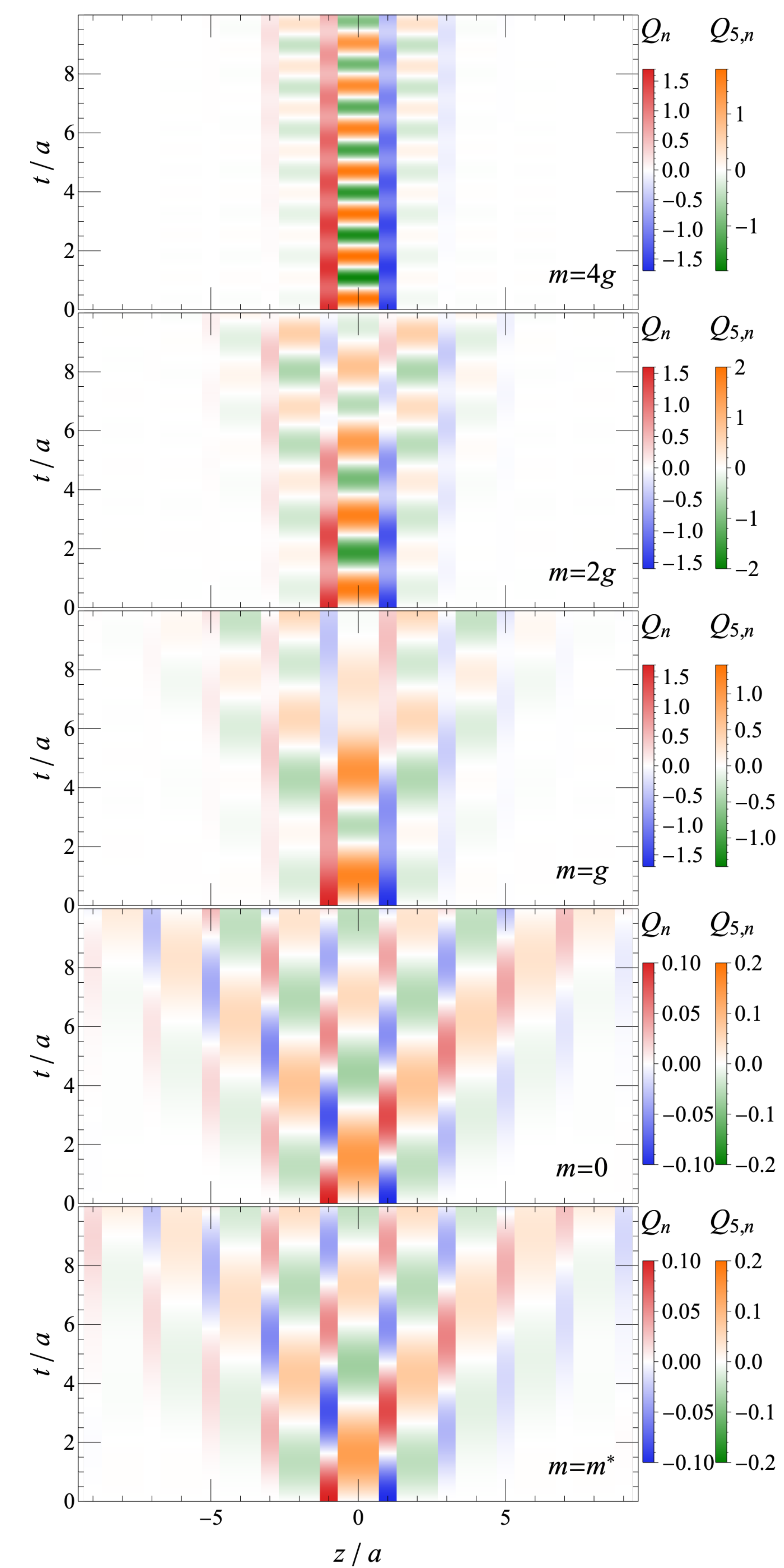}
    \caption{(from top to bottom) Propagation of local electric charge and electric current for $m=4g$, $m=2g$, $m=g$, $m=0$, and $m=m^*=-0.2g$.  Expectation values of the vacuum states have been subtracted. (Same as Fig.1 but with different initial condition. Will explain how the initial condition was prepared when we discuss.)}
    \label{fig:cmw}
\end{figure}

\emph{The quench protocol}. --- 
To study the real-time dynamics of CMW, we first prepare the ground state of the system and then perform a quench by introducing at time $t=0$ an electric dipole at the center of our lattice. The time evolution is then described by the Hamiltonian \eqref{eq:Ham}.

To be more specific, the Hamiltonian~\eqref{eq:Ham} is a high-dimensional, yet sparse, matrix. We first obtain the vacuum state $\ket{0}$ as the Hamiltonian's ground state by exact diagonalization. 
Then at $t=0$ we introduce the electric dipole at the center of our lattice. 
We truncate the Hilbert space by keeping a hundred of the lowest energy levels, and write the state at $t=0$ as
\begin{align}
    \ket{\psi}_{t=0} = \sum_{k=0}^{N_\mathrm{trunc}} c_k \ket{k},
\end{align}
where the superposition coefficients are fixed so that the initial state contains a vector charge dipole on top of the vacuum,
\begin{align}\begin{split}
\bra{\psi} Q_{n} \ket{\psi}_{t=0} =\;& \bra{0} Q_{n} \ket{0} + D\,(\delta_{n,\frac{N}{2}}-\delta_{n,\frac{N}{2}+1}),\\
    \bra{\psi} Q_{5,n} \ket{\psi}_{t=0} =\;& \bra{0} Q_{5,n} \ket{0}.
\end{split}\end{align}
Here, $D$ is the magnitude of electric charges in the dipole, and we choose the coefficients so that $D$ is maximized, within our truncated Hilbert space and on our lattice.
With the initial state prepared, we evolve the quantum state according to $\ket{\psi}_{t} = e^{-iH t} \ket{\psi}_{t=0}$, and measure the local and global observables as a function of time.

We set up the calculation with 20 staggered sites ($N=20$), and take the lattice spacing as $a=0.5/g$. We explore several values of the fermion mass: massive ($m=4g$, $2g$, and $g$), massless ($m=0$), and critical mass of the phase transition ($m=m^*=-0.2g$). For each of these values, we solve the vacuum state and time evolution and measure the local charges. Results are presented in Fig.~\ref{fig:cmw}. Here, we have combined the staggered fermion-anti-fermion pairs to obtain the physical vector charge density, i.e., $Q(z_n) = Q_{2n-1} + Q_{2n}$, whereas the axial charge is defined on the link between two sites, $Q_5(z_{n+1/2}) = 2\,Q_{5,2n}$. The charge conservation requirement~\eqref{eq:charge_conservation}, as manifested in staggered observables, $\partial_t{Q}_{2n-1} = (Q_{5,2n-1} - Q_{5,2n-2})/a$ and similarly for $\partial_t{Q}_{2n}$, automatically lead to the continuity equation\footnote{Note that a physical site contains two lattice sites, therefore it has volume $2a$.}, $\partial_t Q(z_n) = \frac{Q_5(z_{n+1/2})-Q_5(z_{n-1/2})}{2a}$.

At each site, we observe oscillations of the local vector and axial charges, and these oscillations propagate from the middle of the lattice toward its edges forming a light-cone structure. For the massive $m\neq0$ scenario, we observe a stronger damping of the oscillation amplitude, compared to the massless and critical point scenarios. This can be attributed to the non-linear nature of CMW in the massive case.

\emph{Thumper Solution}. ---
In particular, in the massive cases of $m = g$, $2g$, and $4g$, we observe that the oscillation period for vector charge is much longer than that of the axial charge, and the ratio between them increases with fermion mass. This is especially striking in the case of $m=4g$ (see the upper panel of Fig.~\ref{fig:cmw}): the electric dipole is nearly static and does not oscillate at all, whereas the axial charge rapidly oscillates within the dipole. We refer to this solution as a ``thumper". Unfortunately, so far we have not been able to find the corresponding classical solution of ~\eqref{eq:classical_cmw} analytically. 

\begin{figure}[!hbtp]\centering
    \includegraphics[width=0.45\textwidth]{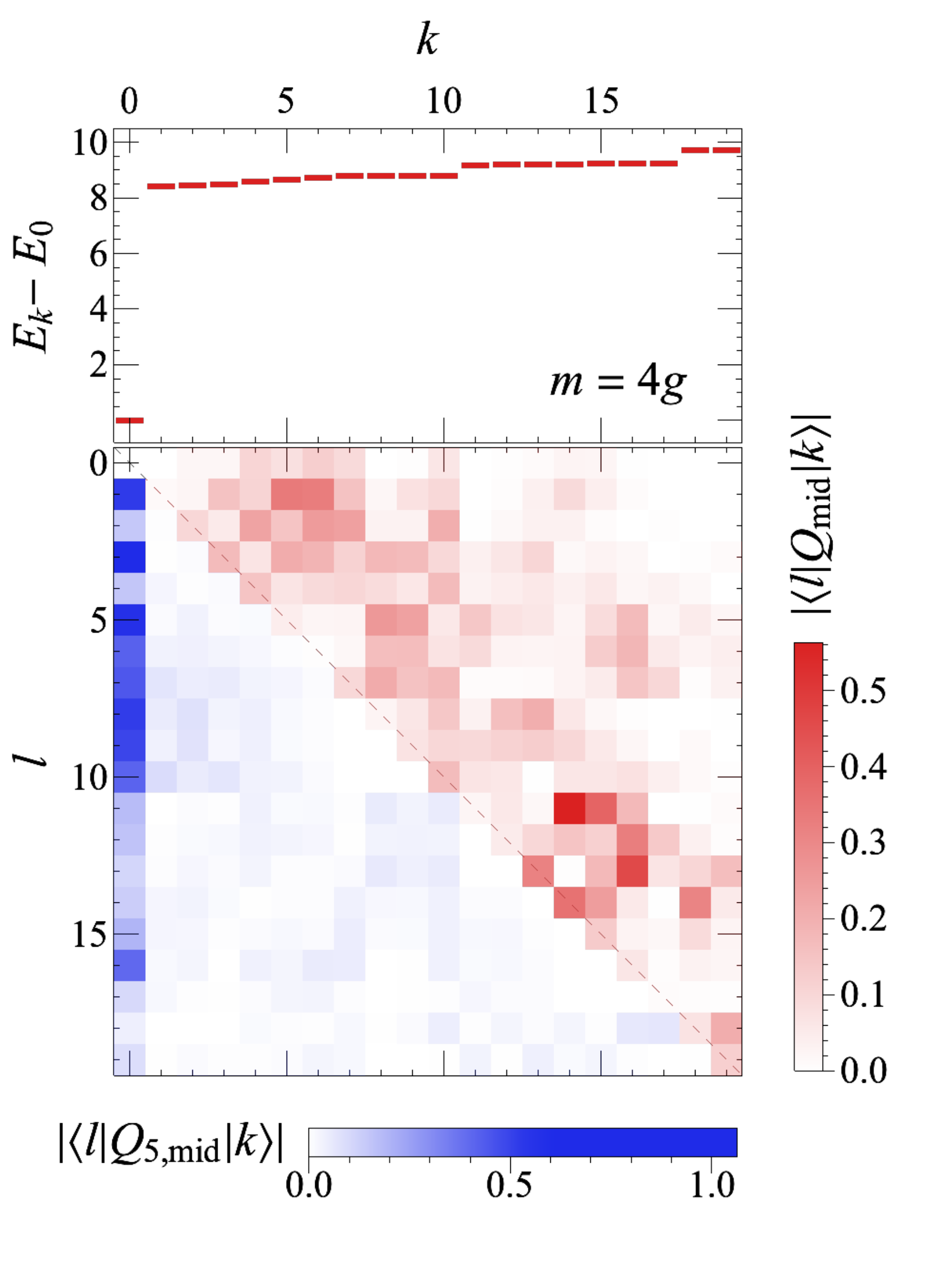}
    \caption{Energy level (upper) and modular of matrix elements (lower) for vector(upper triangle) and axial(lower triangle) charges. 
    Note that $|\bra{l} Q \ket{k}| = |\bra{k} Q \ket{l}|$ and likewise for $Q_5$. Also, $|\bra{k} Q_5 \ket{k}| = 0$. }
    \label{fig:matrix_element}
\end{figure}

To understand the real-time evolution of the vector and axial charges in terms of the eigenstates of Hamiltonian,
\begin{align}
    H \ket{k} = E_k \ket{k}.
\end{align}
we start with the initial state
\begin{align}
    \ket{\Psi(t=0)} = \sum_{k} c_k \ket{k},
\end{align}
and consider the time dependence of an operator $O$ in the Heisenberg picture
\begin{align}
\begin{split}
    O(t) 
    \equiv\;& \bra{\Psi(t)} O \ket{\Psi(t)} \\
    =\;& \sum_{k,l} c_l^*\, c_k\, e^{i(E_l-E_k)t} \bra{l} O \ket{k}.
\end{split}
\end{align}
It is a superposition of different oscillation modes. In each mode, the oscillation frequency is the difference between two energy eigenvalues.
In Fig.~\ref{fig:matrix_element}, we show the energy eigenvalues and  matrix elements for operators $Q_\mathrm{mid} \equiv Q(z_{N/2})$ and $Q_\mathrm{5,mid} \equiv Q_{5}(z_{(N+1)/2})$. While we have extracted 100 lowest energy states in the $Q=0$ subspace, only those with non-vanishing overlap $(|c_k|^2)$ with the initial states are presented in this plot. We note that all excited states are above the ground state energy by $E_k - E_0 \gtrsim 2m$, which correspond to bound states consisting of a fermion--anti-fermion pair. Meanwhile, the energy difference between different excited states is of the order of $\mathcal{O}(g)$. From Fig.~\ref{fig:matrix_element}(lower), it is clear that the axial charge operator is dominated by the excitation between the vacuum and the excited bound states (see leftmost column), whereas the vector charge operator is dominated by scattering between excitations. Therefore, in the massive limit that $2m \gg g$, the oscillation frequency of the axial charge is much greater than that of the vector charge. In the massless limit $(m=0)$ or critical mass $m=m^*$, there is no longer $\sim 2m$ mass gap between the vacuum and the bound states, and the axial and vector charges oscillate with the same frequency.

\emph{Conclusion}. --- We have presented the study of real-time dynamics of chiral magnetic waves (CMWs) in massless and massive Schwinger model using quantum simulations on a classical hardware. For Schwinger model tuned to the conformal critical point ($\theta=\pi$, $m/g \simeq 0.2$) we have observed a gapless CMW propagating with the speed of light. For massless case, we found a gapped CMW corresponding to the familiar non-interacting massive boson representation. 

In the case of a massive Schwinger model ($m/g > 1$), we have uncovered the existence of novel ``thumper" solutions in which the electric charge density oscillates much slower than the axial charge density. In particular, at $m/g =4$ we have observed a nearly static electric dipole with rapid oscillations of chiral charge confined within. Qualitatively, this happens because at large $m/g$ it becomes difficult to break the confining electric string between the charges in the dipole, which prevents the dipole from expanding. On the other hand, the interplay of chiral anomaly and the large fermion mass results in rapid oscillations of chiral charge inside the string. 
The confining electric string thus contains rapid fluctuations of chiral charge. It will be interesting to explore the possible link \cite{Kharzeev:2015xsa,Kharzeev:2015ifa} between the fluctuations of topology and confinement in $(3+1$ dimensions.
It will also be interesting to explore the non-linear CMWs and the ``thumpers" in real systems.

\section*{Acknowledgement}
This work was supported by the U.S. Department of Energy, Office of Science, National Quantum Information Science Research Centers, Co-design Center for Quantum Advantage (C2QA) under Contract No.DE-SC0012704 (KI), and the U.S. Department of Energy, Office of Science, Office of Nuclear Physics, Grants Nos. DE-FG88ER41450 (DK, SS) and DE-SC0012704 (DK).

\appendix
\bibliographystyle{utphys}
\bibliography{ref}

\providecommand{\href}[2]{#2}\begingroup\raggedright\begin{thebibliography}{10}

\bibitem{Kharzeev:2004ey}
D.~Kharzeev, ``{Parity violation in hot QCD: Why it can happen, and how to look
  for it},'' \href{http://dx.doi.org/10.1016/j.physletb.2005.11.075}{{\em Phys.
  Lett. B} {\bfseries 633} (2006) 260--264},
  \href{http://arxiv.org/abs/hep-ph/0406125}{{\ttfamily arXiv:hep-ph/0406125}}.

\bibitem{Kharzeev:2007jp}
D.~E. Kharzeev, L.~D. McLerran, and H.~J. Warringa, ``{The Effects of
  topological charge change in heavy ion collisions: 'Event by event P and CP
  violation'},'' \href{http://dx.doi.org/10.1016/j.nuclphysa.2008.02.298}{{\em
  Nucl. Phys. A} {\bfseries 803} (2008) 227--253},
  \href{http://arxiv.org/abs/0711.0950}{{\ttfamily arXiv:0711.0950 [hep-ph]}}.

\bibitem{Fukushima:2008xe}
K.~Fukushima, D.~E. Kharzeev, and H.~J. Warringa, ``{The Chiral Magnetic
  Effect},'' \href{http://dx.doi.org/10.1103/PhysRevD.78.074033}{{\em Phys.
  Rev. D} {\bfseries 78} (2008) 074033},
  \href{http://arxiv.org/abs/0808.3382}{{\ttfamily arXiv:0808.3382 [hep-ph]}}.

\bibitem{Kharzeev:2013ffa}
D.~E. Kharzeev, ``{The Chiral Magnetic Effect and Anomaly-Induced Transport},''
  \href{http://dx.doi.org/10.1016/j.ppnp.2014.01.002}{{\em Prog. Part. Nucl.
  Phys.} {\bfseries 75} (2014) 133--151},
  \href{http://arxiv.org/abs/1312.3348}{{\ttfamily arXiv:1312.3348 [hep-ph]}}.

\bibitem{Kharzeev:2015znc}
D.~E. Kharzeev, J.~Liao, S.~A. Voloshin, and G.~Wang, ``{Chiral magnetic and
  vortical effects in high-energy nuclear collisions\textemdash{}A status
  report},'' \href{http://dx.doi.org/10.1016/j.ppnp.2016.01.001}{{\em Prog.
  Part. Nucl. Phys.} {\bfseries 88} (2016) 1--28},
  \href{http://arxiv.org/abs/1511.04050}{{\ttfamily arXiv:1511.04050
  [hep-ph]}}.

\bibitem{Landsteiner:2016led}
K.~Landsteiner, ``{Notes on Anomaly Induced Transport},''
  \href{http://dx.doi.org/10.5506/APhysPolB.47.2617}{{\em Acta Phys. Polon. B}
  {\bfseries 47} (2016) 2617},
  \href{http://arxiv.org/abs/1610.04413}{{\ttfamily arXiv:1610.04413
  [hep-th]}}.

\bibitem{Kharzeev:2020jxw}
D.~E. Kharzeev and J.~Liao, ``{Chiral magnetic effect reveals the topology of
  gauge fields in heavy-ion collisions},''
  \href{http://dx.doi.org/10.1038/s42254-020-00254-6}{{\em Nature Rev. Phys.}
  {\bfseries 3} no.~1, (2021) 55--63},
  \href{http://arxiv.org/abs/2102.06623}{{\ttfamily arXiv:2102.06623
  [hep-ph]}}.

\bibitem{Kharzeev:2010gd}
D.~E. Kharzeev and H.-U. Yee, ``{Chiral Magnetic Wave},''
  \href{http://dx.doi.org/10.1103/PhysRevD.83.085007}{{\em Phys. Rev. D}
  {\bfseries 83} (2011) 085007},
  \href{http://arxiv.org/abs/1012.6026}{{\ttfamily arXiv:1012.6026 [hep-th]}}.

\bibitem{Klco:2018kyo}
N.~Klco, E.~F. Dumitrescu, A.~J. McCaskey, T.~D. Morris, R.~C. Pooser, M.~Sanz,
  E.~Solano, P.~Lougovski, and M.~J. Savage, ``{Quantum-classical computation
  of Schwinger model dynamics using quantum computers},''
  \href{http://dx.doi.org/10.1103/PhysRevA.98.032331}{{\em Phys. Rev. A}
  {\bfseries 98} no.~3, (2018) 032331},
  \href{http://arxiv.org/abs/1803.03326}{{\ttfamily arXiv:1803.03326
  [quant-ph]}}.

\bibitem{Butt:2019uul}
N.~Butt, S.~Catterall, Y.~Meurice, R.~Sakai, and J.~Unmuth-Yockey, ``{Tensor
  network formulation of the massless Schwinger model with staggered
  fermions},'' \href{http://dx.doi.org/10.1103/PhysRevD.101.094509}{{\em Phys.
  Rev. D} {\bfseries 101} no.~9, (2020) 094509},
  \href{http://arxiv.org/abs/1911.01285}{{\ttfamily arXiv:1911.01285
  [hep-lat]}}.

\bibitem{Magnifico:2019kyj}
G.~Magnifico, M.~Dalmonte, P.~Facchi, S.~Pascazio, F.~V. Pepe, and
  E.~Ercolessi, ``{Real Time Dynamics and Confinement in the $\mathbb{Z}_{n}$
  Schwinger-Weyl lattice model for 1+1 QED},''
  \href{http://dx.doi.org/10.22331/q-2020-06-15-281}{{\em Quantum} {\bfseries
  4} (2020) 281}, \href{http://arxiv.org/abs/1909.04821}{{\ttfamily
  arXiv:1909.04821 [quant-ph]}}.

\bibitem{Shaw:2020udc}
A.~F. Shaw, P.~Lougovski, J.~R. Stryker, and N.~Wiebe, ``{Quantum Algorithms
  for Simulating the Lattice Schwinger Model},''
  \href{http://dx.doi.org/10.22331/q-2020-08-10-306}{{\em Quantum} {\bfseries
  4} (2020) 306}, \href{http://arxiv.org/abs/2002.11146}{{\ttfamily
  arXiv:2002.11146 [quant-ph]}}.

\bibitem{Kharzeev:2020kgc}
D.~E. Kharzeev and Y.~Kikuchi, ``{Real-time chiral dynamics from a digital
  quantum simulation},''
  \href{http://dx.doi.org/10.1103/PhysRevResearch.2.023342}{{\em Phys. Rev.
  Res.} {\bfseries 2} no.~2, (2020) 023342},
  \href{http://arxiv.org/abs/2001.00698}{{\ttfamily arXiv:2001.00698
  [hep-ph]}}.

\bibitem{Ikeda:2020agk}
K.~Ikeda, D.~E. Kharzeev, and Y.~Kikuchi, ``{Real-time dynamics of Chern-Simons
  fluctuations near a critical point},''
  \href{http://dx.doi.org/10.1103/PhysRevD.103.L071502}{{\em Phys. Rev. D}
  {\bfseries 103} no.~7, (2021) L071502},
  \href{http://arxiv.org/abs/2012.02926}{{\ttfamily arXiv:2012.02926
  [hep-ph]}}.

\bibitem{2023arXiv230111991F}
A.~{Florio}, D.~{Frenklakh}, K.~{Ikeda}, D.~{Kharzeev}, V.~{Korepin}, S.~{Shi},
  and K.~{Yu}, ``{Real-time non-perturbative dynamics of jet production:
  quantum entanglement and vacuum modification},''
  \href{http://dx.doi.org/10.48550/arXiv.2301.11991}{{\em arXiv e-prints}
  (Jan., 2023) arXiv:2301.11991},
  \href{http://arxiv.org/abs/2301.11991}{{\ttfamily arXiv:2301.11991
  [hep-ph]}}.

\bibitem{PhysRevD.107.L071502}
K.~Ikeda, ``Criticality of quantum energy teleportation at phase transition
  points in quantum field theory,''
  \href{http://dx.doi.org/10.1103/PhysRevD.107.L071502}{{\em Phys. Rev. D}
  {\bfseries 107} (Apr, 2023) L071502}.
  \url{https://link.aps.org/doi/10.1103/PhysRevD.107.L071502}.

\bibitem{Ikeda:2023zil}
K.~Ikeda, D.~E. Kharzeev, R.~Meyer, and S.~Shi, ``{Detecting the critical point
  through entanglement in Schwinger model},''
  \href{http://arxiv.org/abs/2305.00996}{{\ttfamily arXiv:2305.00996
  [hep-ph]}}.

\bibitem{Bauer:2022hpo}
C.~W. Bauer {\em et~al.}, ``{Quantum Simulation for High Energy Physics},''
  \href{http://arxiv.org/abs/2204.03381}{{\ttfamily arXiv:2204.03381
  [quant-ph]}}.

\bibitem{Schwinger:1962tp}
J.~S. Schwinger, ``{Gauge Invariance and Mass. 2.},''
  \href{http://dx.doi.org/10.1103/PhysRev.128.2425}{{\em Phys. Rev.} {\bfseries
  128} (1962) 2425--2429}.

\bibitem{Kogut:1974ag}
J.~B. Kogut and L.~Susskind, ``{Hamiltonian Formulation of Wilson's Lattice
  Gauge Theories},'' \href{http://dx.doi.org/10.1103/PhysRevD.11.395}{{\em
  Phys. Rev.} {\bfseries D11} (1975) 395--408}.

\bibitem{Susskind:1976jm}
L.~Susskind, ``{Lattice Fermions},''
  \href{http://dx.doi.org/10.1103/PhysRevD.16.3031}{{\em Phys. Rev.} {\bfseries
  D16} (1977) 3031--3039}.

\bibitem{Jordan:1928wi}
P.~Jordan and E.~P. Wigner, ``{About the Pauli exclusion principle},''
\href{http://dx.doi.org/10.1007/BF01331938}{{\em Z. Phys.} {\bfseries 47}
  (1928) 631--651}.

\bibitem{Kharzeev:2015xsa}
D.~E. Kharzeev and E.~M. Levin, ``{Color Confinement and Screening in the
  $\theta$ Vacuum of QCD},''
  \href{http://dx.doi.org/10.1103/PhysRevLett.114.242001}{{\em Phys. Rev.
  Lett.} {\bfseries 114} no.~24, (2015) 242001},
  \href{http://arxiv.org/abs/1501.04622}{{\ttfamily arXiv:1501.04622
  [hep-ph]}}.

\bibitem{Kharzeev:2015ifa}
D.~E. Kharzeev, ``{Color confinement from fluctuating topology},''
  \href{http://dx.doi.org/10.1142/S0217751X16450238}{{\em Int. J. Mod. Phys. A}
  {\bfseries 31} no.~28n29, (2016) 1645023},
  \href{http://arxiv.org/abs/1509.00465}{{\ttfamily arXiv:1509.00465
  [hep-ph]}}.

\end{thebibliography}\endgroup
\clearpage
\begin{widetext}
\section{Eigenstate analysis of oscillations in vector and axial charges}
For compleness, we present the energy eigenvalues and the matrix elements of vector and axial charge operators for fermion mass $m=2g$, $g$, $0$, and critical mass $m^*$.
\begin{figure*}[!hbtp]\centering
    \includegraphics[width=0.4\textwidth]{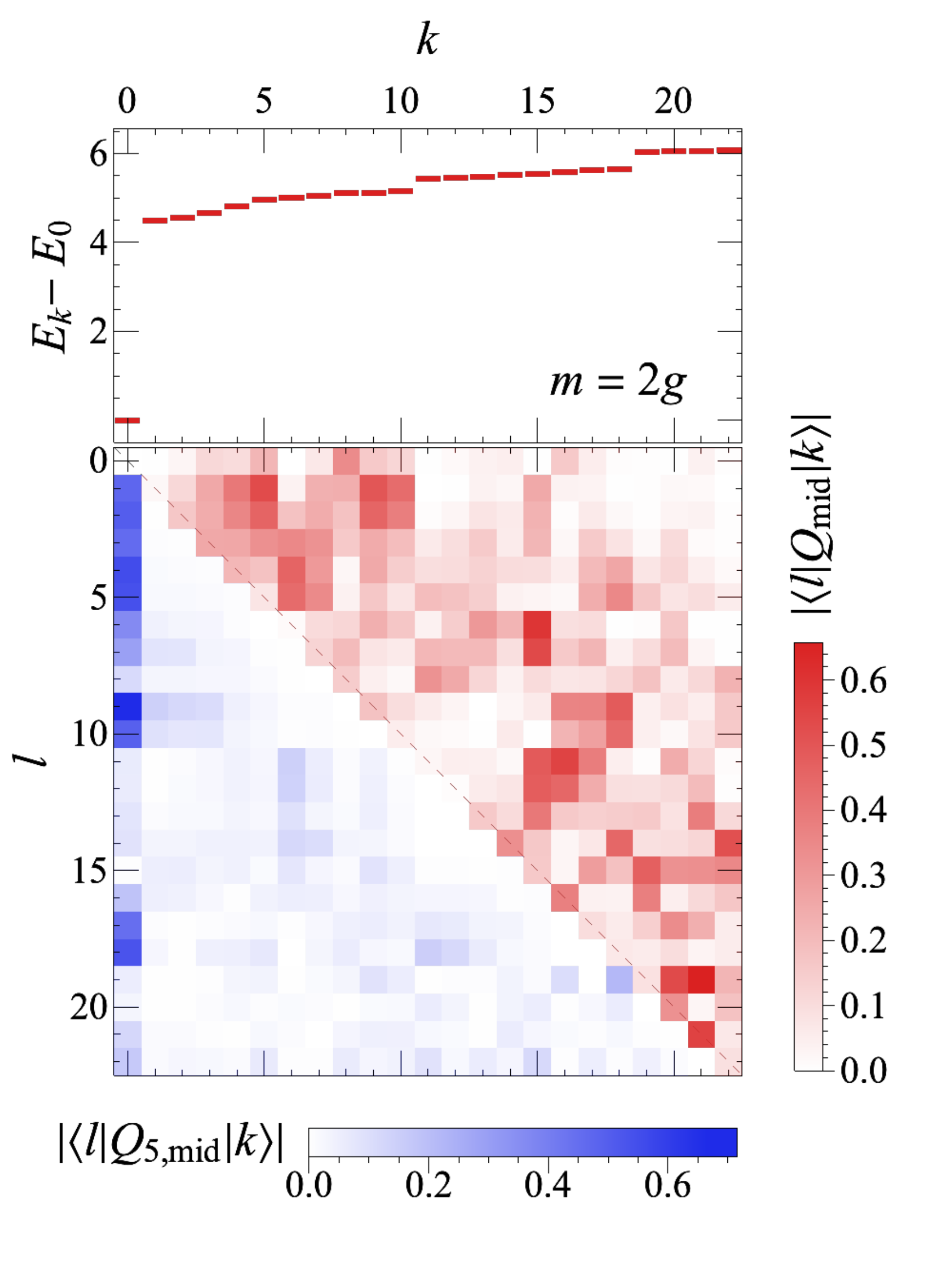}
    \includegraphics[width=0.4\textwidth]{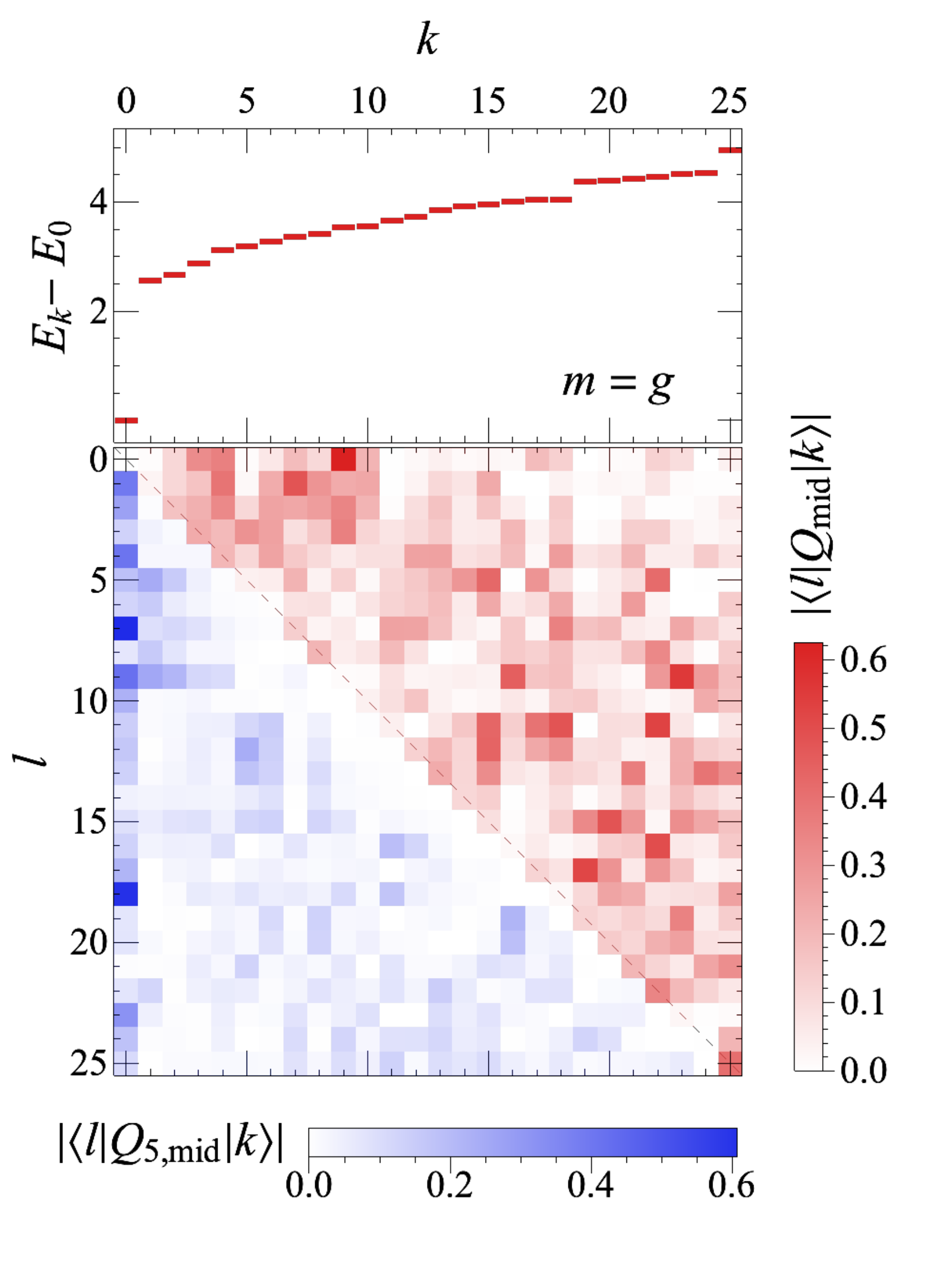}\\
    \includegraphics[width=0.4\textwidth]{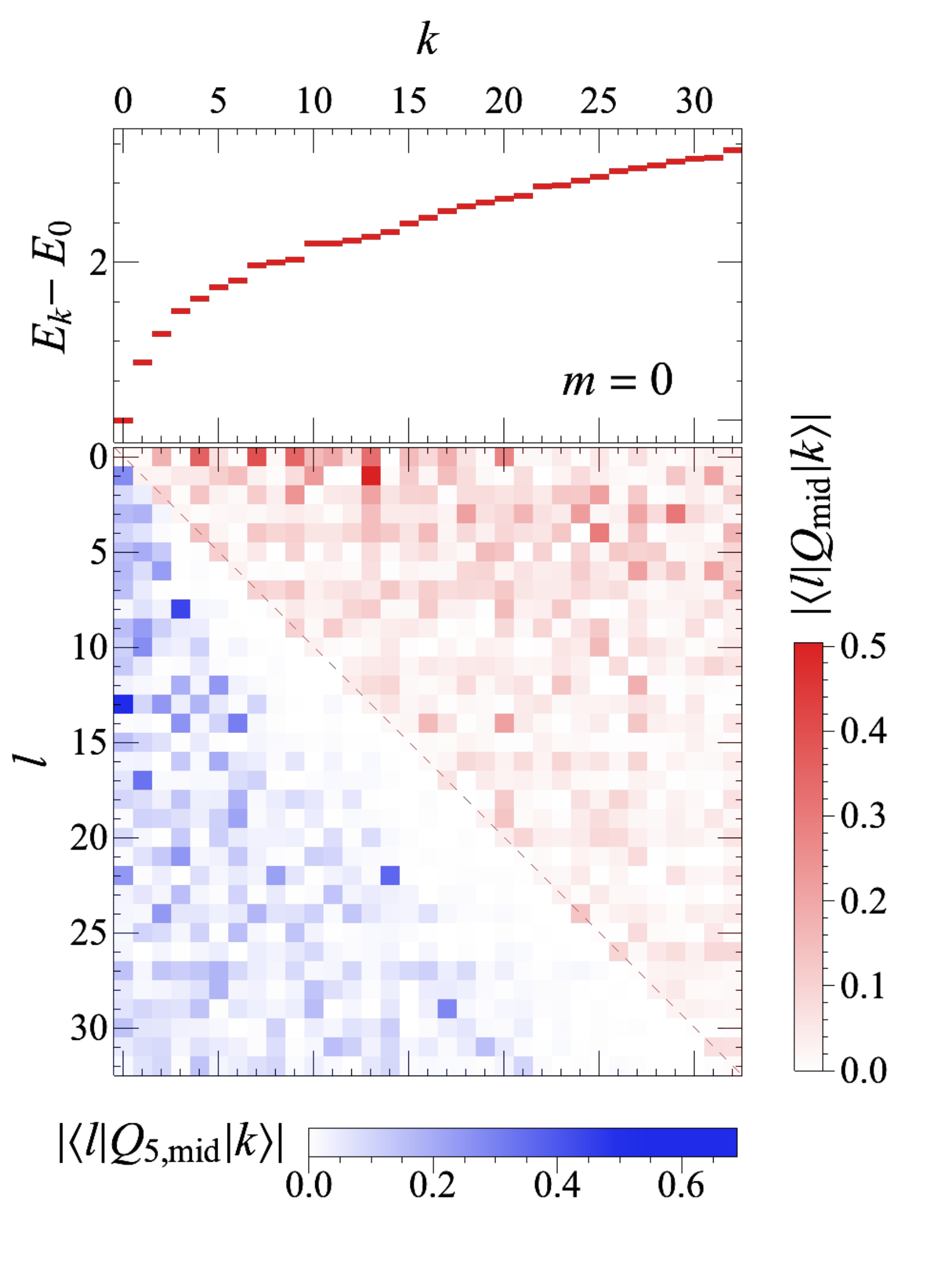}
    \includegraphics[width=0.4\textwidth]{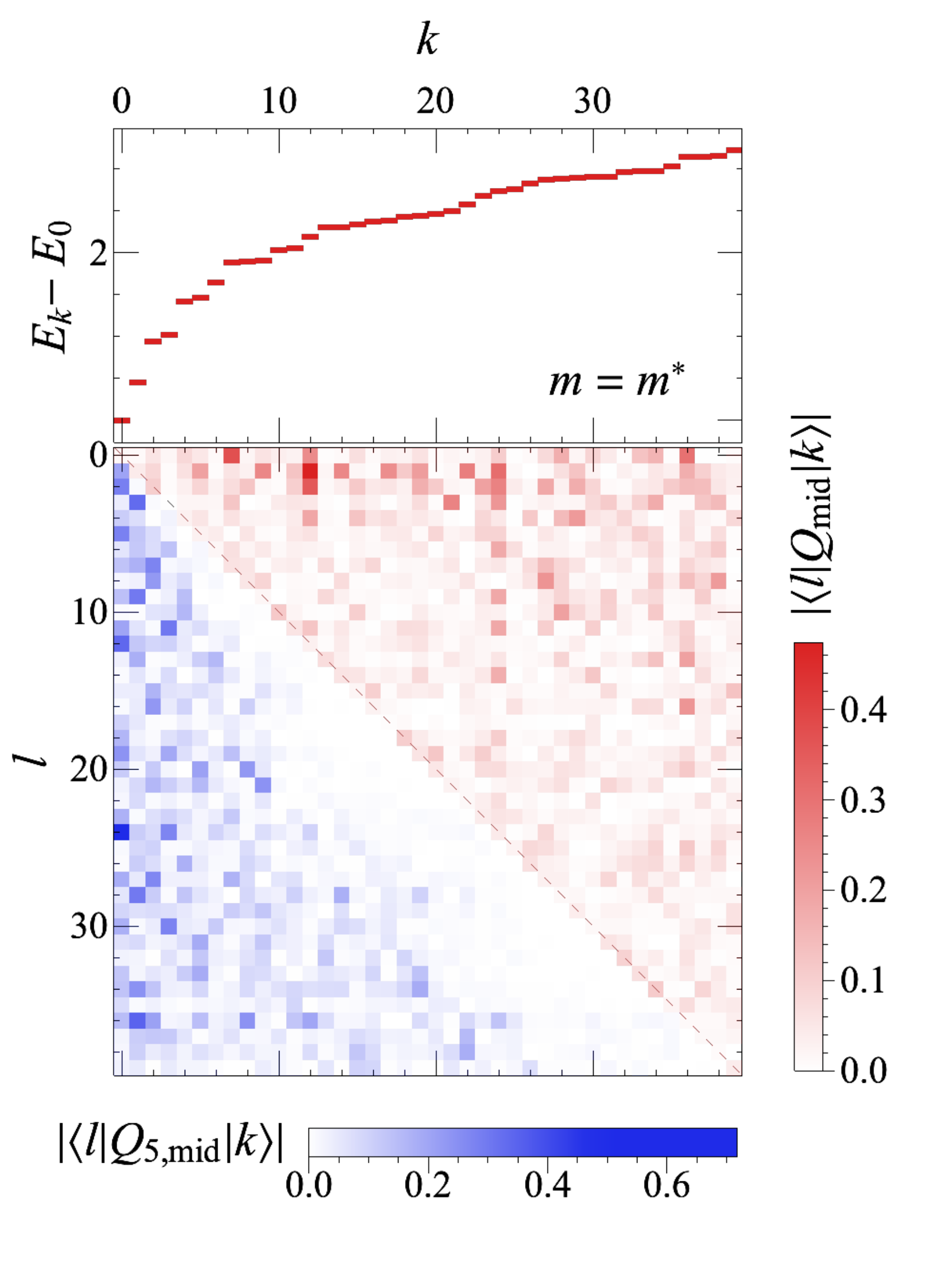}
    \caption{Same as Fig. 2 of the main text but for masses $m=2g$ (upper left), $g$ (upper right), $0$ (lower left), and $m^*$ (lower right). }
    \label{fig:scattering_matrix_2}
\end{figure*}

\end{widetext}
\end{document}